\documentclass{article}

\usepackage{microtype}
\usepackage{graphicx}
\usepackage{booktabs} 
\usepackage{amsmath} 
\usepackage{amssymb} 

\usepackage{hyperref}

\newcommand{\passk}[1]{pass\textasciicircum #1}

\usepackage[accepted]{icml2019}

\begin{document}
\raggedbottom

\twocolumn[
\title{BC-Bench: Evaluating Agentic Engineering in a Domain-Specific Language for ERP}
\date{\vspace{-0.2in}}
\maketitle



\icmlsetsymbol{equal}{*}

\begin{icmlauthorlist}
\icmlauthor{Haoran Sun, Microsoft, haoransun@microsoft.com}{equal}
\icmlauthor{Klaus Marius Hansen, Microsoft, klaus.marius.hansen@microsoft.com}{equal}
\end{icmlauthorlist}

\vspace{0.4in}
\begin{abstract}
Agentic engineering systems have shown strong performance on general-purpose benchmarks, yet their effectiveness in enterprise resource planning (ERP) domain-specific languages (DSLs) remains underexplored. We introduce BC-Bench, a benchmark designed to evaluate agentic engineering on real-world tasks in AL, the DSL for Microsoft Dynamics 365 Business Central. BC-Bench comprises 101 manually curated tasks extracted from two Microsoft-owned production repositories, reflecting authentic ERP development workflows. Adapting the SWE-Bench methodology, we address the unique constraints of the AL ecosystem---including limited public resources and complex environment provisioning. Beyond generating functional code, BC-Bench evaluates test generation and supports multimodal problem statements where visual context is commonly present. We evaluate multiple frontier models across two agent harnesses, utilizing multi-run metrics to account for nondeterminism. In the Bug Fixing category, under our evaluated settings, between-model differences in resolution rate are larger than differences between the two evaluated agent harnesses, and improvements reported on general-purpose benchmarks do not consistently transfer to AL. These results highlight the need for domain-specific evaluation.

\end{abstract}
\vspace{0.4in}
]

\printAffiliationsAndNotice{\icmlEqualContribution} 
\clearpage

\section{Background}

Despite rapid advancements in language model (LM)-based code generation, there is limited work on evaluating agentic engineering workflows for Domain-Specific Languages (DSLs). Existing benchmarks primarily focus on general-purpose programming languages like Python \cite{jimenez_swe-bench_2024, deng_swe-bench_2025}, while benchmarks targeting DSLs remain relatively limited. Results from general-purpose benchmarks do not necessarily transfer to DSL settings \cite{cassano_knowledge_2024}.

Microsoft Dynamics 365 Business Central is ERP software for small and medium-sized businesses. Its business logic is implemented in a DSL called AL. AL is a statically typed language inspired by Pascal. An AL program is made up of objects including pages (UI), codeunits (business logic) and tables (backed by real SQL tables), which are compiled to .NET assemblies that are deployed into a Business Central environment and run against its database. AL tests are themselves codeunits, marked with \texttt{Subtype = Test}, and they require a live Business Central environment to run. This tight coupling between code and a running environment, together with AL conventions around object triggers, record validation, permissions, and domain-specific test libraries, gives AL characteristics not commonly seen in general-purpose languages. Figure~\ref{fig:al-snippets} shows two short examples of AL code.

Evaluating LM-based coding agents on DSLs in the ERP domain presents several practical challenges. First, publicly available resources for dataset construction are limited, as industrial codebases are often proprietary. For example, at the time of writing (February 2026), GitHub search returned about 2.4M repositories for Python and 581K for C\#, compared to about 1.7K for ABAP (an ERP language used in SAP systems) and 338 for AL under the MIT license\footnote{GitHub repository search snapshot for AL with MIT license (accessed February 2026): \href{https://github.com/search?q=language\%3AAL+license\%3AMIT\&type=repositories}{language:AL license: MIT github.com}.} (these time-varying counts are included to illustrate relative scale rather than exact totals). The scarcity of resources necessitates significant manual effort to curate a realistic and diverse dataset. Second, DSL ecosystems rely on domain-specific tooling and infrastructure, making it difficult to adapt existing benchmarking frameworks \cite{joel_survey_2025,yang_swe-bench_2024}.

\begin{figure}[t]
\vskip 0.05in
\begin{center}
\begin{minipage}{0.98\columnwidth}
{\scriptsize
\noindent\textbf{(a) Table object with auto-assigned No.}
\begin{verbatim}
table 18 Customer
{
    fields
    {
        field(1;   "No.";        Code[20]) { }
        field(107; "No. Series"; Code[20])
        {
            TableRelation = "No. Series";
        }
    }

    trigger OnInsert()
    var
        NoSeries: Codeunit "No. Series";
    begin
        if "No." = '' then
            "No." := NoSeries.GetNextNo("No. Series");
    end;
}
\end{verbatim}

\vspace{2pt}

\noindent\textbf{(b) Test codeunit}
\begin{verbatim}
codeunit 134530 "No. Series Tests"
{
    Subtype = Test;

    [Test]
    procedure FailsWhenSeriesRunsOut()
    var
        NoSeries: Codeunit "No. Series";
        NoSeriesCode: Code[20];
    begin
        Initialize();

        asserterror NoSeries.GetNextNo(NoSeriesCode);
        LibraryAssert.ExpectedError(CannotAssignNewErr);
    end;
}
\end{verbatim}
}
\end{minipage}
\end{center}
\caption{Simplified AL examples: a \texttt{Customer} table that auto-assigns its primary key from a No. Series, and a test codeunit that exercises No. Series behavior.}
\label{fig:al-snippets}
\vskip -0.1in
\end{figure}

Business Central's ecosystem is highly extensible and customizable, with thousands of partner-developed applications available on Microsoft Marketplace. As of February 2026, Microsoft Marketplace listed about 8.4K partner-published apps for Business Central\footnote{Microsoft Marketplace partner-app search snapshot for Business Central (accessed February 2026): \href{https://marketplace.microsoft.com/en-us/search/products?product=dynamics-365-business-central\&page=1\&filters=partners}{marketplace.microsoft.com}.} and about 8.8K partner-published apps across all Microsoft 365 products\footnote{Microsoft Marketplace partner-app search snapshot for Microsoft 365 products (accessed February 2026): \href{https://marketplace.microsoft.com/en-us/search/products?page=1\&filters=partners\&product=excel\%3Bofficemetaos\%3Bonenote\%3Bpowerpoint\%3Boutlook\%3Bproject\%3Bsharepoint\%3Bteams\%3Bword\%3Bviva}{marketplace.microsoft.com}.}. While numbers vary with time, this highlights the scale and diversity of real-world AL development. Measuring and improving agentic engineering performance in AL is critical to the development and maintenance of this ecosystem.

We introduce BC-Bench, a benchmark for evaluating coding agents on real-world Business Central/AL engineering tasks, including a dataset of 101 manually curated tasks from two production repositories: a private repository NAV and a public repository BCApps \footnote{\href{https://github.com/microsoft/BCApps}{https://github.com/microsoft/BCApps}}. BC-Bench is open source and publicly available on GitHub \footnote{\href{https://github.com/microsoft/BC-Bench}{https://github.com/microsoft/BC-Bench}}. BC-Bench uses real-world agent harnesses, GitHub Copilot and Claude Code, to closely approximate practical usage by AL engineers. It is designed to support experimentation with different agent configurations, such as custom instructions, tools, and skills. The benchmark runs directly in GitHub Actions without additional implementation, reducing reproducibility issues that arise from minor implementation differences \cite{biderman_lessons_2025}.

This paper makes the following contributions: (1) a benchmark for evaluating coding agents on real-world AL tasks; (2) a manually curated dataset from production repositories; (3) an evaluation harness that enables reproducible experimentation with agent configurations in an AL-specific environment.

\section{Related Work}

Recent work has introduced benchmarks for evaluating LM-based coding agents on real-world software engineering tasks. SWE-Bench \cite{jimenez_swe-bench_2024} pioneered evaluation using real-world Python issues. Extensions include SWE-Bench Multimodal \cite{yang_swe-bench_2024} which incorporates visual elements and expands to additional languages, and SWE-Bench-Live \cite{zhang_swe-bench_2025} which introduced RepoLaunch to simplify the environment setup for different programming languages. Addressing some known limitations of SWE-Bench, SWE-Bench Pro \cite{deng_swe-bench_2025} was introduced to improve task realism and mitigate contamination risks. Terminal-Bench \cite{merrill_terminal-bench_2026} goes beyond academic agent harnesses and uses real-world agent harnesses like Claude Code to better reflect actual developer workflow.

While these benchmarks provide robust evaluation for general-purpose programming languages, they do not capture the constraints and characteristics of DSLs. A survey of code generation for DSLs notes that evaluation in this area faces challenges including scarce data, limited tooling, and domain-specific execution workflows \cite{joel_survey_2025}. Hardware description languages such as Verilog and VHDL provide concrete examples: benchmarks in these languages use domain-specific specifications and verification workflows, and show that LMs still face performance gaps on hardware design tasks \cite{jin_realbench_2025, vijayaraghavan_vhdl-eval_2024}. These efforts underscore the need for DSL-specific evaluation and show that code generation performance depends on the language, tooling, and domain context.

Within ERP systems, similar challenges have been observed in the SAP and ABAP ecosystem, where benchmarking is hindered by limited data availability and specialized infrastructure \cite{wallraven_benchmarking_2026}.

Benchmarking efforts for the AL ecosystem remain limited. AL development involves proprietary codebases, domain-specific abstractions, and tightly coupled business logic that existing benchmarks do not adequately represent. Practitioner-driven efforts indicate growing interest in this space. Notably, CentralGauge \footnote{\href{https://github.com/SShadowS/CentralGauge}{https://github.com/SShadowS/CentralGauge}} is a community benchmark inspired by HumanEval \cite{chen_evaluating_2021} and MBPP \cite{austin_program_2021}, and evaluates LM proficiency across different aspects of the AL language, such as data modeling and event patterns. However, it does not target repository-level engineering tasks and real-world problems. These limitations motivate the need for benchmarks that capture real-world engineering workflows in AL environments, which we address with BC-Bench.

\section{Methods}

BC-Bench is inspired by SWE-Bench and adapted to the Business Central/AL ecosystem.

\subsection{Agent Harnesses}

An agent harness is the system that orchestrates model interaction with the task environment, including prompting, tool use, file edits, and submission handling. We evaluate two production harnesses: GitHub Copilot and Claude Code.

Because agent harnesses evolve rapidly, often with multiple releases per week, we track harness-version differences explicitly, as described in Section~\ref{sec:version-strategy}.

\subsection{Dataset Construction}

The dataset construction consists of a five-stage pipeline.

\textbf{Stage 1: Repository selection.} We manually review merged pull requests (PRs) from two Microsoft-maintained AL repositories that contain the core Business Central application logic. NAV is an internal repository hosted on Azure DevOps, while BCApps is a public repository hosted on GitHub. To the best of our knowledge, there are no actively maintained public AL repositories with comprehensive test suites outside of Microsoft-maintained projects.

\textbf{Stage 2: Attribute-based filtering.} We filter PRs to include only those that fix a single bug and introduce at least one test. This ensures that generated patches can be evaluated by executing the associated test suite.

\textbf{Stage 3: Manual filtering.} We manually review bug descriptions (i.e., problem statements) and exclude instances lacking sufficient context for resolution. We further inspect the associated tests to remove cases where tests depend directly on the implementation of the fix (e.g., tests verifying newly introduced error messages via exact string matching).

\textbf{Stage 4: Bug-PR pair extraction.} We adopt the SWE-Bench collection pipeline to extract bug-PR pairs that satisfy the previous criteria, with several modifications to support the AL ecosystem: (1) \textit{Multi-project mapping}, which identifies the application paths relevant to a given task within a monolithic codebase; (2) \textit{Release-based environment anchoring}, which replaces ephemeral CI builds with stable minor release versions (e.g., BC 26.4; see Appendix~\ref{app:bc-release-lifecycle}) to ensure long-term reproducibility; and (3) \textit{Base-commit alignment}, which selects a release-compatible base commit to resolve inconsistencies between PRs and anchored environments. Following SWE-Bench terminology, we refer to the human-authored fix in the merged PR as the \textit{gold patch}. Additionally, following SWE-Bench Multimodal \cite{yang_swe-bench_2024}, we support screenshots in problem statements to capture the visual nature of ERP bug reports.

\textbf{Stage 5: Execution filtering.} We execute an automated validation pipeline to ensure that each task satisfies correctness constraints: tests that fail prior to the fix must pass after applying the patch, while tests that pass prior to the fix must continue to pass. This pipeline is executed upon task creation and re-run weekly to detect potential regressions due to changes in underlying dependencies or release versions.

Table~\ref{tab:dataset_stats} summarizes the key characteristics of the final BC-Bench dataset for both Bug Fixing and Test Generation categories, and Table~\ref{tab:area_distribution} shows the distribution of tasks across functional areas. Most tasks are from BaseApp, a monolithic application with more than 2 million lines of code (LoC) and broad functional coverage, making it one of the most challenging applications in the ecosystem. The remaining tasks originate from specialized applications, such as Shopify integration, that depend on BaseApp.

The dataset spans a diverse set of functional areas, with a concentration in core business domains such as Inventory, Finance, and Sales. Taken together, these properties make BC-Bench a realistic testbed for AL engineering tasks.

\begin{table}[b]
\vskip 0.05in
\begin{center}
\begin{small}
\begin{sc}
\begin{tabular}{lr}
\toprule
\textbf{Metric} & \textbf{Value} \\
\midrule
Total tasks & 101 \\
BaseApp tasks & 85 \\
Other App tasks & 16 \\
Avg. files modified (gold patch) & 1.3 \\
Avg. gold patch LoC & 18.9 \\
Median files modified (gold patch) & 1.0 \\
Median gold patch LoC & 9.0 \\
Tasks with images & 67 \\
\bottomrule
\end{tabular}
\end{sc}
\end{small}
\end{center}
\caption{Key statistics of the BC-Bench dataset. Averages and medians are computed over all tasks.}
\label{tab:dataset_stats}
\vskip 0.1in
\end{table}

\begin{table}[t]
\vskip 0.05in
\begin{center}
\begin{small}
\begin{sc}
\begin{tabular}{lr}
\toprule
\textbf{Area} & \textbf{Count} \\
\midrule
Inventory & 21 \\
Finance & 19 \\
Sales & 12 \\
Project & 8 \\
Shopify & 7 \\
Manufacturing & 5 \\
Warehouse & 5 \\
CRM & 3 \\
Service & 3 \\
Other & 18 \\
\bottomrule
\end{tabular}
\end{sc}
\end{small}
\end{center}
\caption{Distribution of BC-Bench tasks by area. Areas with count less than two are grouped into Other.}
\label{tab:area_distribution}
\vskip 0.1in
\end{table}

\subsection{Category}

We define two evaluation categories using the same dataset.

\textbf{Bug Fixing.} Following SWE-Bench, each trial provides the agent with a problem statement and a codebase snapshot at a specified base commit. The agent is tasked with generating a functional patch that resolves the issue. When present, images referenced in the problem statement are copied into the testbed and linked via relative paths.

\textbf{Test Generation.} We extend the SWE-Bench setup by introducing a Test Generation category using the same dataset. In addition to the problem statement and codebase, the gold patch is applied as unstaged changes. The agent is then tasked with generating a new test that reproduces the issue. The created test should fail on the base commit and pass after applying the gold patch.

\subsection{Evaluation Metrics}

BC-Bench focuses on the following metrics, all aggregated over five independent runs per task to account for the stochasticity of language models. This choice balances statistical reliability with practical evaluation constraints (e.g.\ cost), informed by preliminary experiments with ten runs on an earlier version of the dataset containing 55 tasks.

\textbf{Mean resolution rate with a 95\% confidence interval.} Because LM agents are stochastic, we run the full dataset five times independently. For Bug Fixing, a task is considered resolved if the generated patch builds successfully and all evaluation tests pass; for Test Generation, a task is considered resolved if the created test fails on the base commit and passes after applying the gold patch. For each run, we compute the resolution rate over all tasks, and we report the mean of these five run-level resolution rates. We estimate the 95\% confidence interval by bootstrapping the five run-level resolution rates with 10,000 resamples, using SciPy's BCa method; this interval reflects run-to-run variability due to agent stochasticity on a fixed benchmark task set.

\textbf{\passk{k}}. To measure consistency across repeated runs, we report \passk{k} \cite{yao_tau-bench_2024}. For a task with $n$ total runs and $c$ successful runs, \passk{k} is the probability that a set of $k$ runs all succeed, averaged across tasks:
\[
\text{pass\string^k}
=
\mathbb{E}_{\mathrm{task}}
\left[
\frac{\binom{c}{k}}{\binom{n}{k}}
\right].
\]
A task contributes $0$ when it has fewer than $k$ successful runs. We perform five runs per task and report \passk{5}. In this setting, \passk{5} equals $1$ for a task only if all five runs succeed, and $0$ otherwise, so the aggregate \passk{5} is the fraction of tasks solved successfully in all five runs.

We do not report pass@k, because it emphasizes eventual success after repeated retries, which does not match our target usage setting. In practice, software engineers care more about whether an agent works reliably when asked than whether one out of many retries eventually succeeds.

\textbf{Duration.} We report duration as wall-clock agent execution time, averaged across task runs and then across the five independent evaluation runs. It covers model calls and tool uses, but excludes post-submission benchmark validation. We include it because latency directly affects the usefulness of agent-based workflows in real-world software engineering.

\subsection{Version Strategy}
\label{sec:version-strategy}

BC-Bench adopts a semantic versioning strategy to track changes that may affect evaluation results. Updates to components such as the dataset, execution environment, and agent harnesses are versioned to ensure reproducibility and comparability across experiments while avoiding the cost of re-running every model for every small update. This is particularly important in LM-based evaluation, where minor implementation differences can significantly affect performance \cite{biderman_lessons_2025}. A detailed version-strategy change log is provided in Appendix~\ref{app:version-strategy}.

\subsection{Evaluation Environment}

All evaluations are conducted using GitHub Actions on a standardized Microsoft internal self-hosted runner. The execution environment is provisioned using BcContainerHelper\footnote{\href{https://github.com/microsoft/navcontainerhelper}{https://github.com/microsoft/navcontainerhelper}}, with Docker containers corresponding to each task's specified release version.

The execution environment is used to run tests and verify the outputs of agent harnesses, but is not directly accessible to agents during task solving. In the default setting, agents interact with the AL codebase as plain text, without access to the underlying development environment. This reflects the current tooling reality: in headless environments, AL tooling is not natively available to coding agents.

In selected experiments, we provide agent harnesses with access to AL MCP, a tooling framework designed to address this limitation by exposing parts of the development environment (e.g., compilation) to agents. BC-Bench allows us to evaluate the impact of such tooling on agent performance.

There is no explicit turn or cost limit imposed on agent harnesses. Each trial is instead subject to a wall-clock timeout of 30 minutes, increased to 60 minutes starting in version 0.4. Trials are evaluated only when they produce a valid submission. Runs that fail to produce a submission, including cases of missing output, non-termination, or infrastructure instability, are treated as workflow failures rather than recorded agent failures. These workflow failures are re-run until the workflow complete successfully. To prevent blocked executions from stalling the pipeline, we additionally enforce a workflow-level timeout that aborts runs that do not terminate cleanly.

\section{Results}

\begin{table*}[t]
\vskip 0.05in
\begin{center}
\begin{small}
\begin{sc}
\begin{tabular}{llllll}
\toprule
\textbf{Agent} & \textbf{Model} & \textbf{Mean with 95\% CI} & \textbf{\passk{5}} & \textbf{Duration} & \textbf{Version} \\
\midrule
Claude Code & claude-opus-4.6 & 68.5\% (65.7--71.3\%) & 49.5\% & 284s & 0.2.0 \\
GitHub Copilot & claude-opus-4.6 & 65.1\% (62.6--67.6\%) & 50.5\% & 314s & 0.2.0 \\
GitHub Copilot & gpt-5.2-codex & 60.8\% (59.4--62.2\%) & 49.5\% & 196s & 0.2.2 \\
GitHub Copilot & claude-opus-4.5 & 59.8\% (58.3--61.3\%) & 38.6\% & 172s & 0.2.0 \\
GitHub Copilot & claude-opus-4.5 & 58.4\% (56.6--60.2\%) & 38.6\% & 165s & 0.1.0 \\
Claude Code & claude-opus-4.5 & 57.4\% (55.4--59.4\%) & 31.7\% & 205s & 0.1.0 \\
GitHub Copilot & gpt-5.3-codex & 55.8\% (54.3--57.3\%) & 37.6\% & 106s & 0.2.1 \\
GitHub Copilot & gpt-5.1-codex-max & 53.7\% (51.7--56.8\%) & 36.6\% & 229s & 0.2.2 \\
GitHub Copilot & gpt-4.1 & 16.6\% (15.6--17.2\%) & 5.0\% & 256s & 0.2.2 \\
\bottomrule
\end{tabular}
\end{sc}
\end{small}
\end{center}
\caption{Bug Fixing performance across agent and model configurations on BC-Bench.}
\label{tab:bugfix_results}
\vskip 0.1in
\end{table*}

\begin{table*}[t]
\vskip 0.05in
\begin{center}
\begin{small}
\begin{sc}
\begin{tabular}{llllll}
\toprule
\textbf{Agent} & \textbf{Model} & \textbf{Mean with 95\% CI} & \textbf{\passk{5}} & \textbf{Duration} & \textbf{Version} \\
\midrule
GitHub Copilot & claude-opus-4.6 & 60.4\% (58.8--62.0\%) & 37.6\% & 469s & 0.2.0 \\
GitHub Copilot & claude-opus-4.5 & 45.5\% (43.0--48.0\%) & 20.8\% & 169s & 0.1.0 \\
GitHub Copilot & gpt-5.3-codex & 45.3\% (42.5--48.1\%) & 20.8\% & 155s & 0.2.2 \\
GitHub Copilot & gpt-5.2-codex & 44.0\% (40.8--48.5\%) & 16.8\% & 291s & 0.2.2 \\
\bottomrule
\end{tabular}
\end{sc}
\end{small}
\end{center}
\caption{Test Generation performance across agent and model configurations on BC-Bench.}
\label{tab:test_generation_results}
\vskip 0.1in
\end{table*}

We present results on BC-Bench, with metrics aggregated over five independent runs per configuration. Because benchmark versions differ across configurations, not all comparisons are direct. Comparisons within the same version support stronger conclusions, while cross-version comparisons should be interpreted more cautiously.

Under matched conditions (GitHub Copilot, version 0.2.0, Bug Fixing) in Table~\ref{tab:bugfix_results}, claude-opus-4.6 outperforms claude-opus-4.5, improving mean resolution rate by 5.3 percentage points and \passk{5} by 11.9 percentage points.

Outside the controlled comparisons, claude-opus-4.6 has the highest \passk{5} and mean resolution rate in the reported runs. Because the benchmark version differs across rows, this should not be read as a definitive ranking. The model is also slower than competing systems, pointing to a trade-off between accuracy and latency. Similar patterns appear in Table~\ref{tab:test_generation_results}.

The earlier model GPT-4.1 exhibits a substantial performance gap compared to recent SOTA models, with approximately 4 times lower mean resolution rate and 10 times lower \passk{5}, highlighting the rapid progress of language models over the past year.

\subsection{Model vs Agent Harness Effect}

For each matched comparison, we evaluate both configurations on the same 101 tasks with five runs per task. For each task, we compute the difference in success counts between the configurations, and use a two-sided exact sign-flip permutation test to assess whether the paired differences consistently favor one configuration over the other. Under the null hypothesis, neither configuration has a systematic advantage.

For the Bug Fixing category, we compare Claude Code and GitHub Copilot under the same model and benchmark version in Table~\ref{tab:harness_comparison}; neither comparison reaches the 0.05 significance threshold. We then fix the harness and benchmark version while varying the model in Table~\ref{tab:model_comparison}; both model comparisons are statistically significant at this threshold.

\begin{table}[t]
\vskip 0.05in
\begin{center}
\begin{small}
\resizebox{\columnwidth}{!}{%
\begin{tabular}{llll}
\toprule
\textbf{Fixed model/version} & \textbf{Setting A} & \textbf{Setting B} & \textbf{$p$-value} \\
\midrule
claude-opus-4.5, v0.1.0 & GitHub Copilot 58.4\% & Claude Code 57.4\% & 0.728 \\
claude-opus-4.6, v0.2.0 & Claude Code 68.5\% & GitHub Copilot 65.1\% & 0.075 \\
\bottomrule
\end{tabular}
}
\end{small}
\end{center}
\caption{Matched harness comparisons on Bug Fixing with model and benchmark version fixed. Values are mean resolution rates.}
\label{tab:harness_comparison}
\vskip 0.1in
\end{table}

\begin{table}[t]
\vskip 0.05in
\begin{center}
\begin{small}
\resizebox{\columnwidth}{!}{%
\begin{tabular}{llll}
\toprule
\textbf{Fixed harness/version} & \textbf{Setting A} & \textbf{Setting B} & \textbf{$p$-value} \\
\midrule
GitHub Copilot, v0.2.0 & claude-opus-4.6 65.1\% & claude-opus-4.5 59.8\% & 0.026 \\
GitHub Copilot, v0.2.2 & gpt-5.2-codex 60.8\% & gpt-5.1-codex-max 53.7\% & 0.019 \\
\bottomrule
\end{tabular}
}
\end{small}
\end{center}
\caption{Matched model comparisons on Bug Fixing with harness and benchmark version fixed. Values are mean resolution rates.}
\label{tab:model_comparison}
\vskip 0.1in
\end{table}

These results suggest that, in our evaluated Bug Fixing settings, model choice appears to matter more than the choice between these two agent harnesses. However, this is based on only four matched comparisons and may not generalize; future work should extend this analysis to additional agent harnesses (e.g., Codex CLI) and to other categories such as Test Generation.

\subsection{Cross-Benchmark Comparison}

Model performance patterns on BC-Bench are broadly consistent with those reported on general-purpose benchmarks such as SWE-Bench, though this comparison should be interpreted cautiously. One notable difference appears in Test Generation: with GitHub Copilot and benchmark version fixed, Table~\ref{tab:cross_benchmark_test_generation} shows no statistically significant advantage for GPT-5.3-codex over GPT-5.2-codex, despite GPT-5.3-codex showing improvements on SWE-Bench Pro.

\begin{table}[t]
\vskip 0.05in
\begin{center}
\begin{small}
\resizebox{\columnwidth}{!}{%
\begin{tabular}{llll}
\toprule
\textbf{Fixed setting} & \textbf{Setting A} & \textbf{Setting B} & \textbf{$p$-value} \\
\midrule
GitHub Copilot, v0.2.2 & gpt-5.3-codex 45.3\% & gpt-5.2-codex 44.0\% & 0.672 \\
\bottomrule
\end{tabular}
}
\end{small}
\end{center}
\caption{Matched comparisons on Test Generation with harness and benchmark version fixed. Values are mean resolution rates.}
\label{tab:cross_benchmark_test_generation}
\vskip 0.1in
\end{table}

This result highlights the importance of domain-specific benchmarks, as improvements observed on general-purpose programming tasks do not necessarily transfer to DSLs such as AL.

\subsection{Factors Affecting Accuracy}

In this section, we analyze how different task characteristics affect agent performance, using results from GitHub Copilot with two representative models.

\textbf{Patch Complexity.} We use the number of files changed and LoC in the gold patch as proxies for task complexity. Table~\ref{tab:resolution_by_files} shows a substantial drop in accuracy when tasks require modifying more than one file, with decreases of over 20 percentage points across both models. A similar degradation is observed in Table~\ref{tab:resolution_by_loc}, where accuracy drops by more than 25 percentage points for patches exceeding 10 LoC. These results suggest that current SOTA models still struggle with more complex, multi-file changes.

\begin{table}[t]
\vskip 0.05in
\begin{center}
\begin{small}
\begin{sc}
\begin{tabular}{llll}
\toprule
\textbf{Files Changed} & \textbf{gpt-5.2} & \textbf{opus-4.6} & \textbf{Tasks} \\
\midrule
1  & 65.1\% & 70.2\% & 82 \\
2+ & 42.1\% & 43.2\% & 19 \\
\bottomrule
\end{tabular}
\end{sc}
\end{small}
\end{center}
\caption{Mean resolution rate by files changed in the gold patch.}
\label{tab:resolution_by_files}
\vskip 0.1in
\end{table}

\begin{table}[t]
\vskip 0.05in
\begin{center}
\begin{small}
\begin{sc}
\begin{tabular}{llll}
\toprule
\textbf{LoC Changed} & \textbf{gpt-5.2} & \textbf{opus-4.6} & \textbf{Tasks} \\
\midrule
1--10  & 75.6\% & 78.5\% & 54 \\
11--25 & 38.3\% & 49.6\% & 23 \\
26+    & 49.2\% & 50.0\% & 24 \\
\bottomrule
\end{tabular}
\end{sc}
\end{small}
\end{center}
\caption{Mean resolution rate by LoC changed in the gold patch.}
\label{tab:resolution_by_loc}
\vskip 0.1in
\end{table}

\textbf{Functional Area.} Table~\ref{tab:resolution_by_area} groups tasks by assigned functional area and shows that performance varies across areas. Shopify and Warehouse tasks have lower resolution rates for both representative models, while GPT-5.2 performs better on Manufacturing tasks. Because some areas contain few tasks, these results should be read as diagnostic rather than definitive. This suggests that domain familiarity may affect resolution rate, in addition to task complexity.

\begin{table}[t]
\vskip 0.05in
\begin{center}
\begin{small}
\begin{sc}
\begin{tabular}{llll}
\toprule
\textbf{Area} & \textbf{gpt-5.2} & \textbf{opus-4.6} & \textbf{Tasks} \\
\midrule
Inventory     & 65.7\% & 70.5\% & 21 \\
Finance       & 51.6\% & 62.1\% & 19 \\
Sales         & 65.0\% & 70.0\% & 12 \\
Project       & 62.5\% & 70.0\% & 8 \\
Shopify       & 48.6\% & 54.3\% & 7 \\
Manufacturing & 88.0\% & 64.0\% & 5 \\
Warehouse     & 40.0\% & 56.0\% & 5 \\
Other         & 63.3\% & 64.2\% & 24 \\
\bottomrule
\end{tabular}
\end{sc}
\end{small}
\end{center}
\caption{Mean resolution rate by functional area. Areas with fewer than five tasks are grouped into Other.}
\label{tab:resolution_by_area}
\vskip 0.1in
\end{table}

\textbf{Tasks with images.} Visual elements in bug descriptions are common in the Business Central ecosystem, particularly in reproduction steps. Table~\ref{tab:resolution_by_image_count} shows that tasks with more images exhibit higher resolution rates. We hypothesize that tasks with more images also include more detailed descriptions, helping the agent regardless of whether the images are processed. We find that image count correlates with problem statement length, suggesting that textual richness, rather than visual content itself, may be the primary driver of improved performance. Future work should investigate whether images are necessary for solving these tasks \cite{yang_swe-bench_2024}.

\begin{table}[t]
\vskip 0.05in
\begin{center}
\begin{small}
\begin{sc}
\begin{tabular}{lllll}
\toprule
\textbf{Image} & \textbf{gpt-5.2} & \textbf{opus-4.6} & \textbf{Tasks} & \textbf{Chars} \\
\midrule
0   & 53.5\% & 57.1\% & 34 & 1,514 \\
1--5 & 56.7\% & 60.0\% & 30 & 1,720 \\
6+  & 70.8\% & 76.8\% & 37 & 2,118 \\
\bottomrule
\end{tabular}
\end{sc}
\end{small}
\end{center}
\caption{Mean resolution rate by number of images in the problem statement. Chars denotes the average character count of the problem statement.}
\label{tab:resolution_by_image_count}
\vskip 0.1in
\end{table}

\subsection{Failure Mode Analysis}
\label{sec:failure-modes}

We classify failed trials from a representative strong configuration in Table~\ref{tab:bugfix_results}: GitHub Copilot with claude-opus-4.6 on Bug Fixing v0.2.0. This setting has 505 trials (5 runs $\times$ 101 tasks), including 176 failures across 50 unique tasks.

We first identify \emph{Timeout}, where the agent does not produce a submission before the trial limit, and \emph{Build Failure}, where the submitted patch does not build. For the remaining failures, we use the overlap between the generated patch and the gold patch as a deterministic proxy for localization, meaning whether the agent found the right place to edit: \emph{Incorrect File} edits none of the gold-patch files, \emph{Incorrect Region} edits a gold-patch file but does not overlap any region changed by the gold patch, and \emph{Wrong Solution} overlaps a region changed by the gold patch but still fails.

\begin{table}[t]
\vskip 0.05in
\begin{center}
\begin{small}
\begin{sc}
\begin{tabular}{lrr}
\toprule
\textbf{Failure Mode} & \textbf{Trials} & \textbf{\%} \\
\midrule
Timeout                             &   4 &  2.3 \\
Build Failure                       &  10 &  5.7 \\
Incorrect File                      &  52 & 29.5 \\
Incorrect Region                    &  32 & 18.2 \\
Wrong Solution                      &  78 & 44.3 \\
\midrule
Total                               & 176 & 100  \\
\bottomrule
\end{tabular}
\end{sc}
\end{small}
\end{center}
\caption{Failure modes for GitHub Copilot with claude-opus-4.6 on Bug Fixing v0.2.0 (176 failed trials of 505).}
\label{tab:failure_modes}
\vskip 0.1in
\end{table}

Table~\ref{tab:failure_modes} shows that build failures are uncommon (5.7\%). Most failed trials produce a buildable patch, so compiler feedback remains useful but would not address most failures on its own. Incorrect File and Incorrect Region account for 47.7\% of failures, indicating that finding the right place to edit is a major failure source. This motivates evaluating coding agents with Language Server Protocol (LSP)\footnote{\href{https://microsoft.github.io/language-server-protocol/}{https://microsoft.github.io/language-server-protocol/}} tools such as find references, go-to definition, and symbol search. The largest single bucket is Wrong Solution (44.3\%), where the patch reaches the right region but still implements the wrong fix.

We manually labeled the 22 unique Wrong Solution tasks (78 failed trials). Table~\ref{tab:wrong_solution_manual_labels} summarizes the resulting labels, with task-level labels in Appendix~\ref{app:wrong-solution-audit}. The labels separate the type of semantic miss after the patch has reached the relevant code region: \emph{Incomplete Fix} means the patch leaves an edge case, validation path, or propagation step unresolved; \emph{Incorrect Business Logic} means the patch edits relevant AL code but implements the wrong functional behavior, such as an incorrect condition or calculation; and \emph{Wrong AL/API Usage} means the patch uses the wrong AL construct, trigger, or API. These labels suggest different improvement directions: test execution and edge-case exploration for incomplete fixes, better retrieval of related AL control and data flows for incorrect business logic, and stronger API guidance for AL/API misuse.

\begin{table}[t]
\vskip 0.05in
\begin{center}
\begin{small}
\begin{sc}
\begin{tabular}{lrr}
\toprule
\textbf{Manual Label} & \textbf{Tasks} & \textbf{Failed Trials} \\
\midrule
Incomplete Fix          & 10 & 36 \\
Incorrect Business Logic & 10 & 34 \\
Wrong AL/API Usage      &  2 &  8 \\
\midrule
Total                   & 22 & 78 \\
\bottomrule
\end{tabular}
\end{sc}
\end{small}
\end{center}
\caption{Task-level manual labels for Wrong Solution failures.}
\label{tab:wrong_solution_manual_labels}
\vskip 0.1in
\end{table}

\section{Discussion}

We introduce BC-Bench, a benchmark for evaluating coding agents on real-world AL engineering tasks in the Business Central ecosystem. BC-Bench is intended as a practical testbed, not only a leaderboard, helping guide future work on tools, skills, and agent configurations. For example, Business Central engineering teams use BC-Bench to prioritize tooling investments for coding agents. Our work shows both the importance and the practical challenges of benchmarking DSLs.

In the Bug Fixing category, we observe larger between-model differences than between the two evaluated agent harnesses, and the matched harness comparisons show no statistically significant effect. Model performance patterns on BC-Bench are broadly consistent with general-purpose benchmarks, but we observe meaningful deviations---most notably that GPT-5.3-codex does not outperform its predecessor on BC-Bench despite gains on SWE-Bench Pro. This suggests that improvements on general-purpose programming tasks do not reliably transfer to DSLs. Among task characteristics, gold patch complexity (measured by number of files changed and LoC) is a strong predictor of difficulty, and resolution rates also vary across functional areas.

Our failure analysis shows that unsuccessful trials are dominated by localization errors and semantically wrong solutions rather than build failures, suggesting that improvements should target repository navigation, context retrieval, and test feedback. These observations yield concrete hypotheses for controlled BC-Bench ablations.

\subsection{Iterative refinement}

During BC-Bench development, we encountered issues that affect  evaluation reliability. For example, manual analysis of agent failures on an earlier version revealed underspecified problem statements and unreachable links. These experiences show that real-world software engineering benchmarks require ongoing maintenance, not just one-time construction.

\subsection{Limitations}

\textbf{Limited Dataset Diversity.} The dataset consists of 101 tasks from Microsoft-owned repositories, which do not represent the full diversity of AL development. The included applications are broadly applicable across customers and the partner ecosystem, whereas many partner-developed AL applications target specific industries, regions, or integrations. All tasks are single bug fixes; scenarios such as feature development, upgrades, localization, and performance tuning remain unexplored. These gaps motivate the \hyperref[sec:dataset-evolution]{Dataset Evolution} direction discussed in Future Work.

\textbf{Limited Representation of Real-Development Environment.} By default, agents interact with the codebase as plain text without AL-specific tooling such as compilation, unlike how engineers work in  Visual Studio Code.

\textbf{Dependency on Tests.} For the Bug Fixing category, BC-Bench uses test execution to verify the correctness of the coding agents' output; however, the collected tests do not necessarily guarantee the correctness of the solution.

\subsection{Future Work}

\textbf{Alternative Evaluation Metrics.} Functional correctness, as measured by passing tests, is not sufficient to characterize a high-quality fix. Future work should explore review-oriented evaluation, such as code review, to assess best practices, security risks, and variability in performance and cost metrics such as duration and token consumption.

\textbf{Experiments.} Building on the failure-mode analysis, a useful next step is to run controlled experiments for AL-specific tooling, including AL MCP and AL LSP, both developed by Microsoft. These experiments can evaluate capabilities such as compilation, symbol search, and go-to definition. Whether these interventions yield meaningful gains is an empirical question that BC-Bench enables answering.

\phantomsection\label{sec:dataset-evolution}\textbf{Dataset Evolution.} The current dataset consists of 101 manually curated, static single-bug tasks, which introduces a risk of contamination after release. Future work should explore continuously updated or dynamically generated benchmarks to mitigate this risk and improve long-term validity.

\section{Acknowledgements}
We thank Jiawen Sun for contributions to dataset collection; Angus Taylor for discussions on statistical methodology, including bootstrap confidence intervals; Thaddeus Loke, Esben Nyhuus Kristoffersen, and Darrick Joo for valuable technical guidance and feedback; Alexander Holstrup for infrastructure support; and Joost Bulsink for supporting the public release of BC-Bench. We also thank the anonymous reviewers for their constructive feedback and suggestions, which helped improve the quality and clarity of this paper.

\clearpage

\bibliographystyle{ieeetr}
\bibliography{bcbench}

\clearpage

\appendix
\section{Version Anchoring and Base-Commit Alignment}
\label{app:bc-release-lifecycle}

Business Central versions follow a \textit{major.minor} scheme (e.g., 26.4). Major versions are released biannually and introduce new features and platform-level changes, while minor releases provide incremental updates such as bug fixes and stability improvements.

In BC-Bench, we anchor each task to a specific minor release rather than an ephemeral CI build, whose availability is short-lived. This design improves long-term reproducibility when benchmarks are re-executed.

Anchoring to a minor release introduces a divergence from the original development history: the base commit used for evaluation may differ from the base commit of the corresponding PR. This occurs because the selected release may include changes that were not present at the time the PR was merged, potentially introducing incompatibilities. To address this, we perform \textit{base-commit alignment} by selecting an earlier, compatible commit within the release, excluding commits that introduce incompatibilities with the task.

One caveat is that hotfix backports may introduce inconsistencies in the evaluation environment. To mitigate potential drift, our execution filtering pipeline is re-run weekly to ensure that all tasks remain valid under the anchored environments.

\section{Version History}
\label{app:version-strategy}

\href{https://www.nuget.org/packages/Microsoft.Dynamics.BusinessCentral.Development.Tools}{altool} is a command-line tool used for compiling and packaging AL extensions for Business Central, designed for CI/CD pipelines. It contains the AL MCP server.

\begin{itemize}
\item \textbf{Version 0.1.0} \hfill (2026-01-26)\\
\textbf{Summary:} Successfully curated 101 tasks for both Bug Fixing and Test Generation categories. All tasks passed the five stages of dataset construction.\\
\textbf{Versions:}
\begin{itemize}
\item GitHub Copilot CLI: 0.0.382
\item Claude Code: 2.0.76
\item altool: 17.0.30.49729-beta
\end{itemize}

\item \textbf{Version 0.2.0} \hfill (2026-02-07)\\
\textbf{Summary:} After manual review of the dataset and comparison of the gold patch against agent outputs, clarified problem statements, fixed unreachable links, and made the prompt more precise so that agents ignore localization changes.\\
\textbf{Versions:}
\begin{itemize}
\item GitHub Copilot CLI: 0.0.406
\item Claude Code: 2.1.37
\item altool: 17.0.30.49729-beta
\end{itemize}

\item \textbf{Version 0.2.1} \hfill (2026-02-18)\\
\textbf{Summary:} Updated GitHub Copilot CLI to include support for gpt-5.3-codex.\\
\textbf{Versions:}
\begin{itemize}
\item GitHub Copilot CLI: 0.0.409
\item Claude Code: 2.1.37
\item altool: 17.0.30.49729-beta
\end{itemize}

\item \textbf{Version 0.2.2} \hfill (2026-02-19)\\
\textbf{Summary:} Updated GitHub Copilot CLI and Claude Code to include support for claude-sonnet-4.6.\\
\textbf{Versions:}
\begin{itemize}
\item GitHub Copilot CLI: 0.0.411
\item Claude Code: 2.1.47
\item altool: 17.0.30.49729-beta
\end{itemize}

\item \textbf{Version 0.3.0} \hfill (2026-02-27)\\
\textbf{Summary:} Updated the Test Generation prompt to state more explicitly that the agent is expected to create a new test case.\\
\textbf{Versions:}
\begin{itemize}
\item GitHub Copilot CLI: 0.0.411
\item Claude Code: 2.1.47
\item altool: 17.0.30.49729-beta
\end{itemize}

\item \textbf{Version 0.3.1} \hfill (2026-03-11)\\
\textbf{Summary:} Updated GitHub Copilot CLI and refreshed the list of allowed models.\\
\textbf{Versions:}
\begin{itemize}
\item GitHub Copilot CLI: 1.0.2
\item Claude Code: 2.1.47
\item altool: 17.0.30.49729-beta
\end{itemize}

\item \textbf{Version 0.3.2} \hfill (2026-03-14)\\
\textbf{Summary:} Updated altool to a newer version that fixes several bugs and adds more tools.\\
\textbf{Versions:}
\begin{itemize}
\item GitHub Copilot CLI: 1.0.2
\item Claude Code: 2.1.47
\item altool: 17.0.33.55542
\end{itemize}

\item \textbf{Version 0.4.0} \hfill (2026-03-19)\\
\textbf{Summary:} Improved the options and settings for Claude Code and GitHub Copilot CLI to prevent internet access. Repository setup now sparse-checks out only app folders, improving clone performance. Fixed two PNG file conversions for a dataset entry. Extended the timeout to accommodate the altool compile tool, and fixed altool integration after the update.\\
\textbf{Versions:}
\begin{itemize}
\item GitHub Copilot CLI: 1.0.2
\item Claude Code: 2.1.47
\item altool: 17.0.33.55542
\end{itemize}

\end{itemize}

\clearpage
\onecolumn
\section{Manual Labeling of Wrong Solution Failures}
\label{app:wrong-solution-audit}

This appendix documents the manual labeling behind Table~\ref{tab:wrong_solution_manual_labels}. The label meanings are defined in Section~\ref{sec:failure-modes}; here we describe the assignment protocol and provide the task-level audit table.

We labeled the 22 unique tasks in the Wrong Solution failure mode rather than each failed trial independently, because repeated runs for the same task often produced related or identical patches. For each task, we inspected the problem statement, gold patch, generated patch variants, and failing test output. The audit uses outputs for GitHub Copilot with claude-opus-4.6 on Bug Fixing v0.2.0 from GitHub Actions runs \texttt{21815911073}, \texttt{21822868049}, \texttt{21832838543}, \texttt{21855870368}, and \texttt{21864728490}. Table~\ref{tab:wrong_solution_task_labels} reports the assigned label, failed-trial count, number of unique generated-patch variants, and duplicate-trial count. Duplicate trials count repeated generated patches for the same task, i.e., failed trials minus unique generated patch variants.

\begin{center}
\refstepcounter{table}\label{tab:wrong_solution_task_labels}
\vskip 0.05in
\begingroup
\setlength{\tabcolsep}{4pt}
\begin{scriptsize}
\begin{tabular}{llrrr}
\toprule
\textbf{Instance ID} & \textbf{Manual Label} & \textbf{Failed Trials} & \textbf{Patch Variants} & \textbf{Duplicate Trials} \\
\midrule
\texttt{microsoft\_\_BCApps-4766} & Incomplete Fix & 5 & 2 & 3 \\
\texttt{microsoft\_\_BCApps-5633} & Incomplete Fix & 2 & 2 & 0 \\
\texttt{microsoftInternal\_\_NAV-176426} & Incorrect Business Logic & 5 & 1 & 4 \\
\texttt{microsoftInternal\_\_NAV-177750} & Incomplete Fix & 5 & 5 & 0 \\
\texttt{microsoftInternal\_\_NAV-181900} & Incorrect Business Logic & 2 & 2 & 0 \\
\texttt{microsoftInternal\_\_NAV-182354} & Incomplete Fix & 5 & 1 & 4 \\
\texttt{microsoftInternal\_\_NAV-183399} & Wrong AL/API Usage & 5 & 2 & 3 \\
\texttt{microsoftInternal\_\_NAV-201169} & Incomplete Fix & 5 & 3 & 2 \\
\texttt{microsoftInternal\_\_NAV-205825} & Incorrect Business Logic & 2 & 2 & 0 \\
\texttt{microsoftInternal\_\_NAV-206135} & Incomplete Fix & 3 & 2 & 1 \\
\texttt{microsoftInternal\_\_NAV-206527} & Incorrect Business Logic & 4 & 4 & 0 \\
\texttt{microsoftInternal\_\_NAV-207236} & Incomplete Fix & 1 & 1 & 0 \\
\texttt{microsoftInternal\_\_NAV-208320} & Wrong AL/API Usage & 3 & 3 & 0 \\
\texttt{microsoftInternal\_\_NAV-208649} & Incorrect Business Logic & 5 & 2 & 3 \\
\texttt{microsoftInternal\_\_NAV-211521} & Incorrect Business Logic & 1 & 1 & 0 \\
\texttt{microsoftInternal\_\_NAV-216918} & Incorrect Business Logic & 5 & 5 & 0 \\
\texttt{microsoftInternal\_\_NAV-218323} & Incorrect Business Logic & 1 & 1 & 0 \\
\texttt{microsoftInternal\_\_NAV-218856} & Incomplete Fix & 5 & 5 & 0 \\
\texttt{microsoftInternal\_\_NAV-220036} & Incorrect Business Logic & 4 & 2 & 2 \\
\texttt{microsoftInternal\_\_NAV-220452} & Incorrect Business Logic & 5 & 5 & 0 \\
\texttt{microsoftInternal\_\_NAV-223493} & Incomplete Fix & 4 & 2 & 2 \\
\texttt{microsoftInternal\_\_NAV-224447} & Incomplete Fix & 1 & 1 & 0 \\
\bottomrule
\end{tabular}
\end{scriptsize}
\endgroup
\vskip 0.05in
{\small \textbf{Table~\thetable.} Task-level manual labels and generated patch duplication for the 22 unique tasks in the Wrong Solution failure mode. Patch variants are exact generated-diff variants for a task.}
\end{center}
\vskip 0.1in

\clearpage
\twocolumn

\end{document}